\documentclass[fleqn,usenatbib]{mnras}

\usepackage{newtxtext,newtxmath}

\usepackage[T1]{fontenc}
\usepackage{ae,aecompl}

\usepackage{graphicx}	
\usepackage{amsmath}	
\usepackage{amssymb}	

\title[Radial \& compact MW's stellar halo]{Cosmological insights into the assembly of the radial and compact stellar halo of the Milky Way}

\author[L. M. Elias et al.]{
Lydia M. Elias,$^{1}$\thanks{E-mail: lydia.elias@email.ucr.edu (LME)}
Laura V. Sales,$^{1}$ 
Amina Helmi$^{2}$
and Lars Hernquist$^{3}$
\\
$^{1}$ Department of Physics and Astronomy, University of California, Riverside, 900 University Ave., Riverside, CA, 92507, USA \\
$^{2}$ Kapteyn Astronomical Institute, University of Groningen,
P.O. Box 800, 9700 AV Groningen, The Netherlands\\
$^{3}$ Harvard-Smithsonian Center for Astrophysics, 60 Garden Street, Cambridge, MA 02138, USA\\
}

\begin{document}
\label{firstpage}
\pagerange{\pageref{firstpage}--\pageref{lastpage}}
\maketitle

\begin{abstract}
Recent studies using Gaia DR2 have identified a massive merger in the history of the Milky Way (MW) whose debris is markedly radial and counterrotating. This event, known as the Gaia-Enceladus/Gaia-Sausage (GE/GS), is also hypothesized to have built the majority of the inner stellar halo. We use the cosmological hydrodynamic simulation Illustris to place this merger in the context of galaxy assembly within $\Lambda$CDM. From $\sim$150 MW analogs, $\sim 80 \%$ have experienced at least one merger of similar mass and infall time as GE/GS. Within this sample, 37 have debris as radial as that of the GE/GS, which we dub the Ancient Radial Mergers (ARMs). Counterrotation is not rare among ARMs, with $43 \%$ having $>40 \%$ of their debris in counterrotating orbits. However, the compactness inferred for the GE/GS debris, given its large $\beta$ and its substantial contribution to the stellar halo, is more difficult to reproduce. The median radius of ARM debris is r$_{*,deb}\simeq 45$kpc, while GE/GS is thought to be mostly contained within $r\sim 30$ kpc. For most MW analogs, a few mergers are required to build the inner stellar halo, and ARM debris only accounts for $\sim 12 \%$ of inner accreted stars. Encouragingly, we find one ARM that is both compact and dominates the inner halo of its central, making it our best GE/GS analog. Interestingly, this merger deposits a significant number of stars (M$_*\simeq1.5 \times 10^9 M_\odot$) in the outer halo, suggesting that an undiscovered section of GE/GS may await detection.
\end{abstract}




\section{Introduction}
\label{sec:intro}

Traces of past accretion events in our Galaxy are imprinted as tidal stellar streams in the outer halo of the MW \citep{Belokurov2006}, where the long dynamical times allow for substructure to remain recognizable for several Gyr. The inner regions of our halo, on the other hand, are characterized by shorter dynamical times, $\leq 100$ Myr, requiring full phase-space information to identify the debris of those once coherent structures \citep{Helmi1999, Helmi1999b}. 
The availability of superb data from Gaia DR2 \citep{Gaia2018}, including positions and velocities for millions of stars, unveiled the remnant of one of the largest accretion events in the history of our Galaxy, the Gaia Enceladus \citep[GE, ][]{Helmi2018} or Gaia Sausage \citep[GS, ][]{Belokurov2018} event.

Since GE and GS were discovered independently in different studies and
it is difficult to establish precise membership criteria, the relation
between the two has not always been clearly discussed in the literature \citep[although see][]{Helmi2020}. It is nonetheless evident that there is a large degree of overlap in the stars belonging to GS and GE \citep{Myeong2019, Matsuno2019, Koppelman2019}. The motion of GE/GS stars is largely eccentric, with radial anisotropy $0.8< \beta<0.9$ \citep{Belokurov2018}, as well as  counterrotating \citep{Helmi2018}, both characteristics enhancing the profile of GE/GS as the result of a past accretion event.


Beyond its kinematics, the GE/GS structure also sticks out because of its high stellar metallicity compared to other halo stars. Cross-correlated data with other surveys such as {\sc sdss}, {\sc apogee} or {\sc h3}, among others, measure a metallicity range -1.7<[Fe/H]<-1 for GE/GS stars \citep{Helmi2018,Belokurov2018,Conroy2019}. This, together with the trend measured for [$\alpha$/Fe] vs. [Fe/H], advocates for a somewhat massive progenitor, with stellar mass estimates placing GE/GS  at infall comparable to the SMC, $5 \times 10^8 \leq \rm M_*/\rm M_\odot  \leq 5 \times 10^9$  \citep{Helmi2018,Vincenzo2019,Myeong2019,Mackereth2019}. The stars in GE/GS are old and can be used to constrain the time of the merger to occur on average $\sim 8$-$10$ Gyr ago (age of the Universe $t \sim 4$-$6$ Gyr). This is assuming that the progenitor stopped forming stars at the time of the merger with the MW, a hypothesis that we briefly explore further in this paper.


If these estimates are correct, the GE/GS event should have had important consequences for the present-day structure of the MW. For instance, the merger with a massive GE/GS would stir up the proto-disk of the MW, perhaps triggering dynamical heating and the posterior build up of the old thick disk in our Galaxy \citep{Helmi2018, Belokurov2019,Haywood2018};  a hypothesis confirmed by cosmological simulations identifying GE/GS analogs \citep{Bignone2019, Grand2020}. These studies also suggest that the gas-rich nature of high-redshift mergers would simultaneously imply an enhancement of the MW's star formation rate as the GE/GS coalesces into the central regions, a prediction for which there is already observational evidence \citep{Kruijssen2019}. Somewhat more surprising is the fact that such a catastrophic encounter occurred at all in the MW, whose dominant and dynamically cold stellar disk would not necessarily convey the idea of a major merger happening at all during the past assembly history of our Galaxy.

Arguably, the most important contribution of GE/GS is to the stellar halo of the MW. While locally -- within a few kpc of the solar neighborhood -- stars belonging to GE/GS completely dominate the stellar halo, its general contribution is less well constrained. There is some consensus that GE/GS is an important contributor to the nearby stellar halo, with estimates in the range $30\%$-$50\%$ within the inner $\sim 25$ kpc of the Galaxy \citep{Lancaster2019,Mackereth2019b}. Moreover, the pile-up of stars in the apocenter of GE/GS may solely be responsible for the ``break'' radius in the 3D density of the MW's stellar halo around $r\sim 25$ kpc. However, only a small fraction of the inferred mass for GE/GS has been positively identified. Where are the rest of the  stars brought in by GE/GS today in our Galaxy?

Numerical simulations indicate that such GE/GS mergers are rare \citep{Mackereth2019, Bignone2019}, especially when requiring that they contribute such a large fraction of the inner halo \citep{Fattahi2019}. Typically, the accreted stellar halos of MW-like galaxies are built from more than one  progenitor \citep{Cooper2010,Pillepich2014,Elias2018,Monachesi2019}, with important contributions from at least a few. Given the suggested rarity of the event, large statistical samples of simulated galaxies are required to shed light on the nature and fate of GE/GS-like events within the $\Lambda$CDM model.

In this paper we take advantage of the large population of MW analogs in the Illustris simulations \citep{Vogelsberger2014a,Vogelsberger2014b} to study the frequency of events comparable to GE/GS in the simulations (Sec.~\ref{sec:merger}); the typical morphology and properties of their remnants (Sec.~\ref{sec:diversity}); and their contribution to the stellar halo, including inner and outer regions (Sec.~\ref{sec:halo}). We conclude with results on the unidentified segment of the GE/GS debris and summarize our results in Sec.~\ref{sec:concl}. 

\section{Numerical Simulations}
\label{sec:sims} 
Illustris is a cosmological hydrodynamical simulation run with the {\sc arepo} code \citep{Springel2010} and covering a cubic volume with $106$ Mpc on a side \citep{Vogelsberger2014a, Vogelsberger2014b, Genel2014}. Illustris cosmological parameters are consistent with the CDM model as determined by WMAP-9: $\Omega_m$=0.2726, $\Omega_b$=0.0456, $\Omega_\Lambda$=0.7274, and H$_0$=70.4 km/s/Mpc \citep{Hinshaw2013}. The project\footnote{https://www.illustris-project.org} includes a suite of simulations run with different numerical resolutions and with/without the inclusion of baryons. In this work, we use data from the largest resolution baryonic run, Illustris-1, featuring a mass per particle of 1.6 and 6.3$\times10^6 M_\odot$ for baryons and dark matter, respectively and a  gravitational softening length 0.7 kpc or better.   

Gravitational forces in {\sc arepo} are calculated via an oct-tree approach while the hydrodynamic equations are solved by means of a finite-volume moving mesh technique \citep{Springel2010,Weinberger2019}. Besides cooling and heating of the gas, a variety of baryonic physical processes are added to the code to track the formation and evolution of galaxies. The main features of the model, including the treatment for heating and cooling, star formation and stellar feedback are described in detail in \citet{Vogelsberger2013, Vogelsberger2014b}. In what follows we briefly summarize the main aspects of the baryonic treatment in Illustris, referring the interested reader to the aforementioned work for more specific information.  

The impact of radiation and reionization is followed via a time-dependent spatially-uniform ultraviolet background according to \citet{Nelson2015}. Gas is allowed to cool to $T \sim 10^4 \; \rm K^\circ$ including H, He and metal cooling lines. Gas above a density threshold $n_H$=0.13 cm$^{-3}$ is put into an effective equation of state modeling a two-phase fluid and is allowed to transform into stars stochastically with $1\%$ efficiency per local dynamical time \citep{Springel2003}. The model adopts a Chabrier initial mass function for stars \citep{Chabrier2003} and follows the subsequent stellar evolution according to {\sc STARBURST99} \citep{Leitherer1999}, keeping track of stellar lifetimes, mass loss and metal production.  

Two main sources of stellar feedback are included: stellar winds and supernova explosions. Supernovae play a major role in shaping the baryonic content of galaxies and are implemented in Illustris in kinetic form, adding 100$\%$ of the available energy due to supernovae as a velocity-scaled wind (that depends on the local dark matter velocity dispersion) and a mass-loading inferred from the available supernova energy. Additionally, all halos with virial mass above $M_{200} = 7\times 10^{10} \rm M_\odot$ are seeded with a supermassive black hole which is allowed to grow and exert feedback on the surrounding interstellar and intergalactic gas according to two feedback modes: quasar (high accretion rate) and radio-mode (low accretion rate) following \citet{Sijacki2015}. We use virial quantity definitions corresponding to $200$ times the critical density of the Universe. 

The {\it subfind} halo finder is used to identify halos and galaxies on the fly in Illustris \citep{Springel2001,Dolag2009}. First, groups are identified using space information via the Friends-of-Friends (FoF) algorithm \citep{Davis1985}. Subsequently, self-gravitating subhalos are identified within these groups, giving rise to a catalog of substructure with assigned dark matter and baryonic content. The object at the center of the gravitational potential of each group is defined as the ``central" galaxy, while all other substructure are considered ``satellites". Following previous work in Illustris, we use a fiducial radius of $r \leq 2 * r_h$ to assign particles to galaxies and compute all ``galaxy" properties (stellar and gas mass, angular momentum, etc.); with $r_h$ defined as the half mass radius of the stars for each object. Particles beyond this radius that are not associated to satellites and are within the virial radius of each central are considered part of the stellar/gaseous halos. 
We use the SUBLINK trees to trace the evolution of each subhalo through the 135 snapshots of the simulation. 

Illustris has been shown to reliably reproduce global properties of the galaxy population such as stellar mass functions, stellar mass-halo mass relations, specific star formation rates, distribution of satellite galaxies, HI column density distribution, color distribution, morphology bimodality, the metallicity-environment relation etc. back to z=7 \citep{Genel2014,Sales2015,Rodriguez-Gomez2015,Snyder2015,Genel2016,Kauffmann2016}, and unusal objects such as shell galaxies \citep{Pop2017, Pop2018}.

\begin{figure}
	\includegraphics[width=\columnwidth]{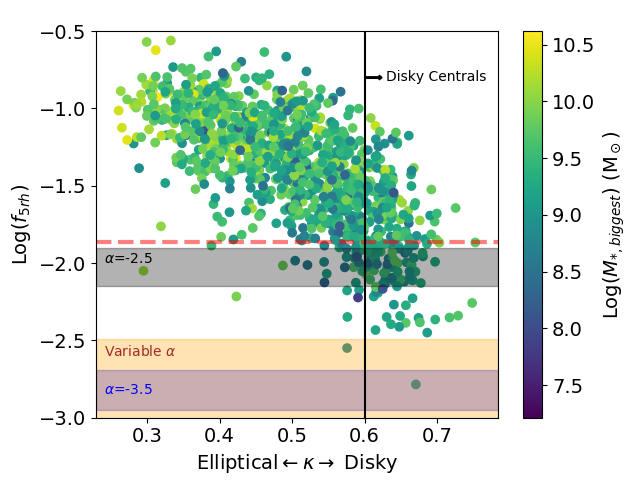}
    \caption{Stellar halo fraction measured for $r>5r_h$ ($f_{5rh}$)  vs.  morphology index $\kappa_{\rm rot}$ for MW-like centrals, colored by the stellar mass of their most massive merger. All galaxies in the sample have experienced at least one merger with a satellite of mass M$_*\geq 10^8 M_\odot$. Disk-dominated centrals (large $\kappa_{\rm rot}$ values) have a lower fraction of mass in their stellar halo component, albeit with significant dispersion. We define a sample of disk-dominated centrals (154 objects) selecting those with $\kappa_{\rm rot} > 0.6$, indicated with the vertical line. The red horizontal line and the shaded regions represent different estimates of $f_{5 rh}$ for the MW (see text for more details). The MW's stellar halo seems consistent with the lowest end of stellar halo fractions in Illustris.} 
    \label{fig:fshvskappa}
\end{figure}

\begin{figure}
	\includegraphics[width=\columnwidth]{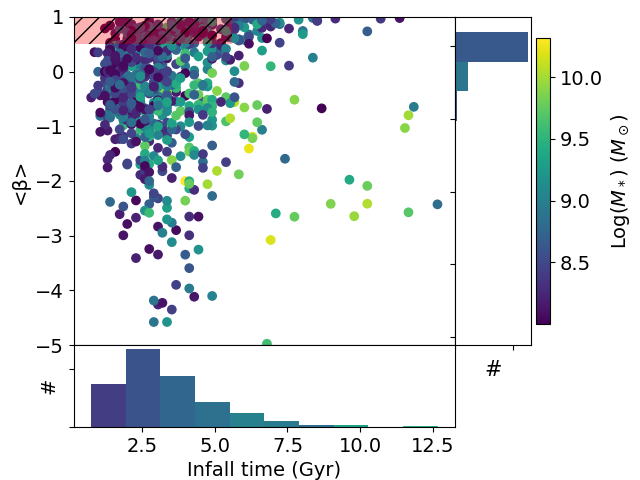}
    \caption{Infall time $t_{\rm inf}$ vs. average orbital anisotropy $< \beta>$ for the stellar debris of all merged satellites in our disk-dominated centrals. Symbols are colored by maximum stellar mass of each satellite. Histograms along both axes show the individual distributions of $t_{\rm inf}$ and $<\beta>$, color-coded by average satellite mass contributing to each bin. 
    Earlier infall times and radially-biased debris are preferred in our sample of disky MW-analogs, although the distributions are wide. Inspired by the GE/GS debris, we study in detail a sample of massive (M$_*>10^{8.5} M_\odot$), early (t$_{inf}<$6 Gyr) and radial ($\beta > 0.5$) mergers,  indicated by the shaded red box. We call them ``Ancient Radial Mergers'', or ARMs for short.}
    \label{fig:betatinf}
\end{figure}
\subsection{Milky Way analogs}
\label{ssec:sample}
We select an initial sample of (central galaxies) MW-analogs in Illustris based on virial mass, $8 \times10^{11}\leq M_{200} \leq 2\times 10^{12} M_\odot$. A statistical analysis of the stellar halos of this sample has already been presented in \citet{Elias2018}. We use the $\kappa_{\rm rot}$ parameter to quantify the stellar morphology of our galaxies. After rotating each galaxy with their total angular momentum pointing along the $z$-direction, $\kappa_{\rm rot}$ compares the total kinetic energy of stars to the energy in co-rotation around the $z$-axis \citep[see ][ for details]{Sales2010,Sales2012}. By construction, large $\kappa_{\rm rot}$ values indicate that a large fraction of the stellar mass is in a disk component supported by rotation. Furthermore, $\kappa_{\rm rot}$ has been shown to correlate well with other techniques to quantify morphology such as dynamical decomposition of galaxies \citep{Abadi2003b, Scannapieco2009}. A total of N=1115 galaxies fall in our virial mass range. 

In Fig.~\ref{fig:fshvskappa} we begin to characterize the stellar halos of this sample of MW-analogs via $f_{5r_h}$, defined as the fraction of stellar mass beyond $5 r_h$ compared to the central galaxy following \citet{Merritt2016}. 
As discussed in \citet{Elias2018}, the fraction of stellar halo correlates with morphology such that spheroid-dominated galaxies have larger fractions of their mass in their extended halos compared to their more disky counterparts. The color-coding in Fig.~\ref{fig:fshvskappa} further indicates that at fixed stellar mass, galaxies with more massive stellar mergers also show a more prominent stellar halo, in agreement with previous results \citep{Monachesi2019,DSouza2018}. 

To select closer analogs to the MW we impose a more stringent cut in disk-like morphology: $\kappa_{\rm rot}$>0.60. This is indicated in Fig.~\ref{fig:fshvskappa} as all points to the right of the vertical black line. With this criterion, our sample of disk-dominated MW analogs consists of $154$ central galaxies. The stellar halo fraction of this sample is lower than considering all galaxies in the virial mass cutoff, which is in agreement with the MW having only a modest stellar halo. The total mass of the MW stellar halo is not well constrained. Different measurement methods as well as definitions of the stellar halo have resulted in significantly varying stellar halo values in the $4.0\times 10^8-1.4\times10^9 \; M_\odot$ range (see e.g. \citet{Deason2019, Carollo2010, Bell2008} for measurements and \citet{Helmi2008,Belokurov2013} for reviews of the literature). 

Given these uncertainties, shaded regions in Fig.~\ref{fig:fshvskappa} indicate several estimates for the fraction of light in the MW stellar halo beyond $5 r_h$ using different assumptions. The dashed red horizontal line corresponds to the quoted estimate for the MW in \citep{Merritt2016}. Assuming the mass of the stellar halo, M$_{SH} = 1.4\pm 0.4 \times 10^{10}  M_\odot$ \citep{Deason2019} and for the disk M$_{disk} = 6.43\pm 0.63 \times 10^9  M_\odot$ \citep{McMillan2011}, the orange shaded region is calculated using a triple power law profile for the stellar halo mass density with slopes measured in  previous studies in the literature: $\alpha= [-2.7,-2.9]$ for $ 0<r\leq25 $ \citep{Xue2015}, $\alpha = [-4.6, -3.8]$ for $25 < r \leq 50$ \citep{Pillepich2014} and $\alpha = [-6.5,-5.5]$ for $50 < r \leq 100 $ \citep{Deason2014}. For simplicity, we also include in black and purple regions the stellar halo fractions assuming a single power law with slopes $\alpha$=-2.5 and $\alpha$=-3.5, respectively. The different $f_{\rm 5 rh}$ estimates tend to populate the lower end of our simulated centrals with disk-like morphology, and hint at an overly-efficient formation of stellar halos in Illustris, as suggested by previous work \citep{DSouza2018}.

\begin{centering}
\begin{figure*}
	\includegraphics[width=2\columnwidth]{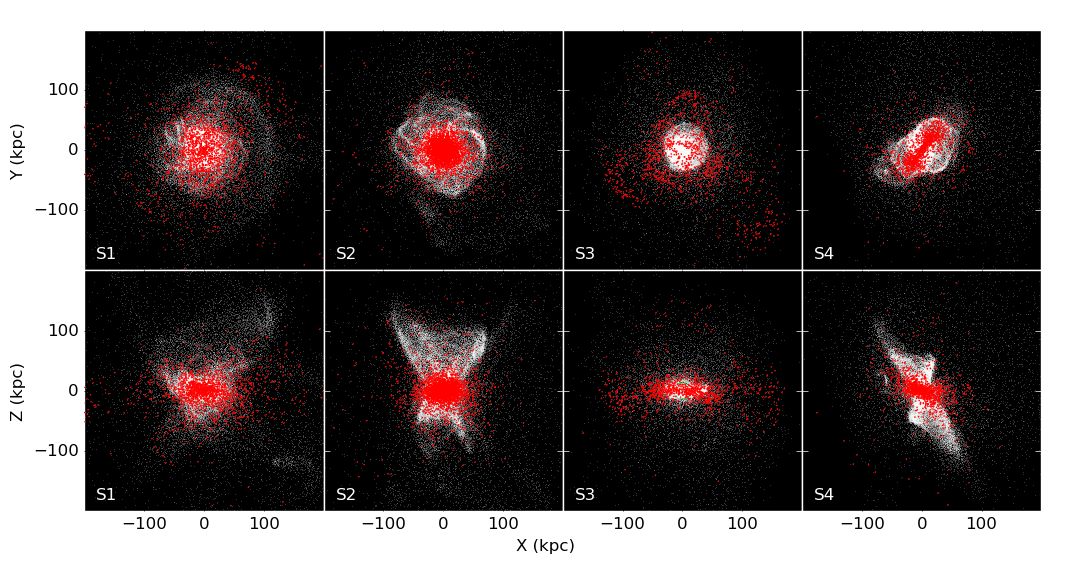}
    \caption{Face-on (top row) and edge-on (bottom row) projections of four representative disk-dominated galaxies with an ancient radial merger (ARM). ARM stellar debris is colored in red. Gas and stellar particles of the central are colored in white. Debris from ARMs can extend out to hundreds of kpc as in columns 1 and 3 (S1 and S3), or be very concentrated, as in columns 2 and 4 (S2 and S4).}
    \label{fig:projs}
\end{figure*}
\end{centering}

\section{GE/GS like mergers for Milky Way analogs} 
\label{sec:merger}

As discussed in Sec.~\ref{sec:intro}, the stellar mass estimates for the GE/GS event are in the range $5 \times 10^8 \leq \rm M_* \leq 5 \times 10^9 \; \rm M_\odot$. We find that in our sample of disky MW analogs,  80$\%$ have experienced at least one merger with a satellite of average stellar mass in this range (M$_*\geq 5\times10^8 M_\odot$). The median stellar mass of the most massive merger experienced by a central in our sample is $M_*\simeq 1.25\times10^9 M_\odot$, suggesting that merger events of mass similar to GE/GS are, in fact, common for Milky Way-like galaxies within $\Lambda$CDM, in agreement with previous findings \citep{Bose2019}. 

The selected merged satellites with mass comparable to GE/GS display a wide range of infall times $t_{\rm inf}$ and orbital anisotropies $\beta$ as shown in Fig.~\ref{fig:betatinf}. Here $\beta$ is computed as 
\begin{equation}
    \beta = 1-\frac{v_T^2}{2v_R^2}
\end{equation}

\noindent
where v$_T$ is the mean velocity in the tangential direction and v$_R$ is the mean velocity in the radial direction of the satellite debris at $z=0$. Infall times are defined as the snapshot at which the galaxy reaches maximum total mass. 

Fig.~\ref{fig:betatinf} also shows the corresponding histograms for both axes, indicating that early infall times ($t_{\rm inf} < 5 \; \rm Gyr$) and rather radial motions ($\beta > 0$) are the norm for this subsample of GE/GS mass-objects. Note that the early infall times obtained for these relatively massive merger events can be considered almost a selection effect in our sample. By selecting disk-dominated morphology for the central host galaxy, we drive a correlation with a low fraction of mass in the stellar halo and also bias high the infall redshift for merger events, as shown in \citet{Elias2018} and expected for our own Galaxy. Massive mergers that occur at late times, shown as light points on the right side of Fig.~\ref{fig:betatinf}, are not as significant as they may first appear since the disk is more massive at those times. Thus, the mass ratio of the satellite to the central is still small enough to result on a disky morphology.

Encouragingly, early infall times are also suggested from observations of  GE/GS, with  estimates of the event placing it at least 8 Gyr ago \citep{Gallart2019,Helmi2018,Mackereth2019} based on a variety of arguments including the age of the youngest stars in the debris. Inspired by this, we refine our sample by defining GE/GS events to have $t_{\rm inf} \leq 5.6$ Gyr (i.e., lookback time $\geq 8$ Gyr). 

Furthermore, Gaia measurements of identified stars belonging to GE/GS also indicate a largely radial orbit, with $0.8<\beta< 0.9$ \citep{Belokurov2018} . We find that such extreme radial orbits are less common in our sample, with only 46 objects (or $\sim$6\%) consistent with such measurement. However, the constraints on $\beta$ may be softened by considering spatial variations along the orbit (as Gaia estimates are rather local to the solar neighborhood while simulated values pertain to the entire debris) and observational errors. Taking this into account, in what follows, we use $\beta > 0.5$ to select the most radial mergers in our disky MW-like galaxies. Satellites in this sample deposited a varying fraction of stars with $0.8<\beta<0.9$, varying from 8\% to 26\%.. 

Our final selection criterion, including cuts in mass ($5 \times 10^8<M_*<5 \times 10^9$), infall times ($t_{\rm inf} < 5.6$ Gyr) and $\beta>0.5$ is highlighted by a red dashed rectangle in Fig.~\ref{fig:betatinf}, resulting in $37$ MW analogs that have experienced a GE/GS-like merger. We refer to this sample of merged satellites as Ancient Radial Mergers (ARMs).

\begin{centering}
\begin{figure*}
	\includegraphics[width=2.25\columnwidth]{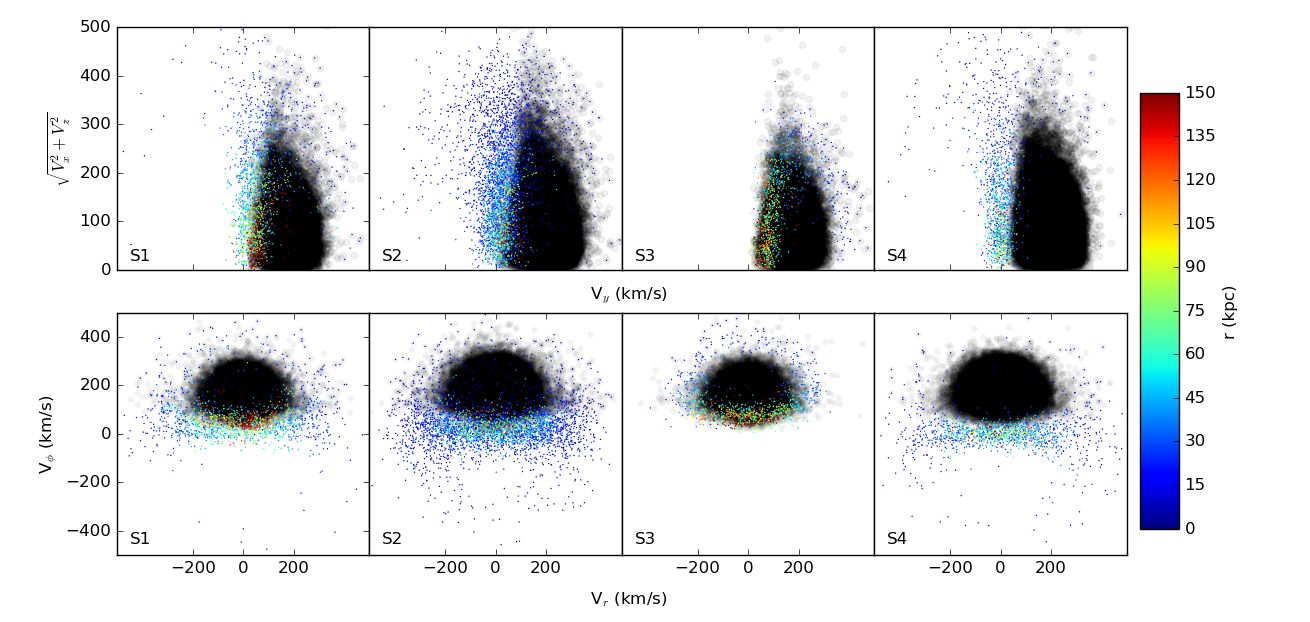}
    \caption{Decomposition of Satellites 1-4 in velocity space. Black points represent stars belonging to the disk, while colored points are stellar debris from S1-S4, colored by galactocentric distance as indicated on the right. Note that all stellar particles in the debris are included in the plot, even beyond 50 kpc, which is different from observational samples that are dominated by stars within $r<50$ kpc. Our sample of massive, early and radial mergers (ARMs) frequently resemble both the Gaia Sausage (GS) {\it and} the Gaia Enceladus (GS), like the case of S2 and S4, suggesting a common origin for these structures. Lack of counter-rotating stars make examples S1 and S3 unlikely matches to GE/GS.}  
    \label{fig:encel}
\end{figure*}
\end{centering}

\section{Diverse morphology and kinematics for the ancient merger debris}
\label{sec:diversity}

 Our highly specialized sample of 37 ARMs demonstrates a considerable degree of diversity in their distribution and kinematics. To ease visualization, we choose four satellite galaxies that represent the variety of the entire sample. Fig.\ref{fig:projs} shows face-on and edge-on projections of these four galaxies, named Satellite 1,2,3, and 4 for simplicity and  hereafter (or S1-S4 for short). White points represent both gas and star particles in the central galaxy, while red points represent the stellar debris of each satellite at $z=0$. For instance, satellites  S1 and S3 have extremely extended debris, with their once-bound stellar particles found today out to several hundred kiloparsecs. By contrast, S2 and S4 have relatively concentrated remnants. Given that the progenitors of these debris have a restricted range of masses, infall times, and orbits, the extreme degree of diversity found on their present-day distribution is somewhat surprising.
 
We can use these examples to shed light on the link between the GS and the GE events. As mentioned in Sec.~\ref{sec:intro}, they are often referred to interchangeably in the literature but given their different identification criteria it is currently unclear if they are the same object. Our sample of satellites has been selected to have extremely radial remnants, which a priori suggests a similarity to the GS. However, in order to keep the analysis general, for now we place no constraints on the velocity with respect to the disk or whether this debris dominates the solar neighborhood, which both the GS and GE debris do. 

Fig.~\ref{fig:encel} makes a direct comparison of the debris of our four selected ARMs in  Fig.~\ref{fig:projs}, using the same identification space of the GE \citep[top, ][]{Helmi2018} and the GS \citep[bottom, ][]{Belokurov2018}. All stars brought in by satellites S1-S4 are colored by their galactocentric distance today. The black points correspond to disk stars associated with our simulated centrals, selected by satisfying our disk criteria: r$_{xy}$<2r$_h$,|z|<3kpc, and circ>0.5. Here r$_{xy}=\sqrt{x^2+y^2}$, r$_h$ is the half light radius, and circularity is defined as:
\begin{equation}
    circ=\frac{j_z}{r\cdot v_{circ}}
\end{equation}
where j$_z$ is the angular momentum in the z-direction (oriented by the disk), r is the 3D radius and v$_{circ}$ is the circular velocity of the particle. 

Satellites S2 and S4 in Fig.~\ref{fig:encel} are good analogs to both GS and GE, demonstrating that it is possible within the $\Lambda$CDM framework to explain this remnant as a result of a single common event. On the other hand, S1 would not be a good candidate to GE but shows certain similarities to GS while S3 results in a debris with no resemblance to either GE or GS. The color gradient in the bottom row of Fig.~\ref{fig:encel} indicates that the accreted stars have larger velocities at smaller galactocentric radii (bluer colors).


\begin{figure}
	\includegraphics[width=\columnwidth]{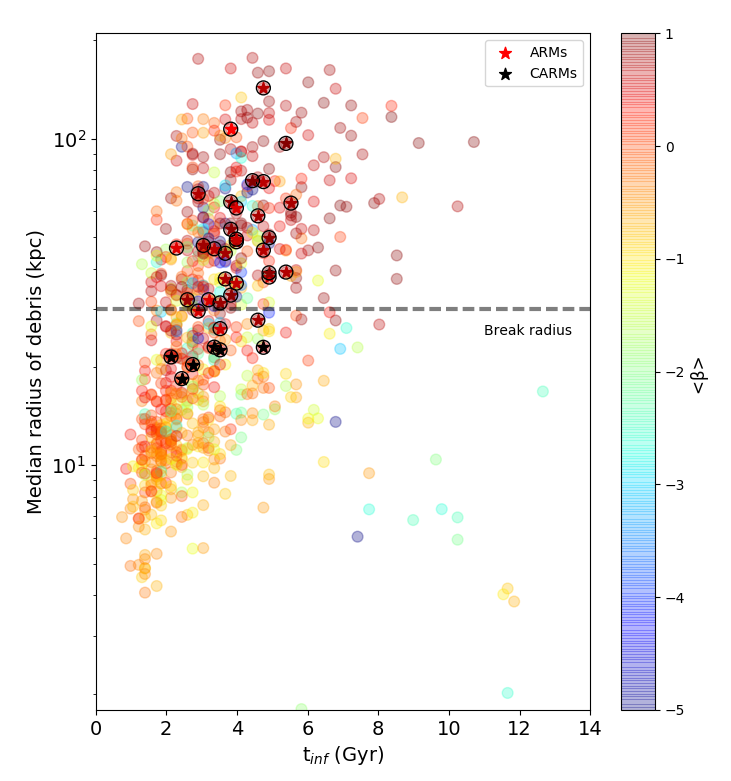}
    \caption{Radius encompassing 50$\%$ of stars in merged satellite debris, $r_{*,\rm deb}$, vs. infall time for our disky MW-like galaxies. Symbols are color-coded according to the median anisotropy of the stars, $<\beta>$, with all mergers shown in semi-transparent circles and ARMs highlighted in solid starred symbols. Naturally, more radial orbits (larger $<\beta>$) corresponds to more extended debris. This is at odds with the compact distribution of stars inferred observationally for GE/GS. The horizontal dashed line indicates the `break' radius in the MW stellar halo, presumably associated with the outer edge of GE/GS. We define a set of Compact ARMs (or CARMs), by requiring additionally $r_{*,\rm deb} < 25$ kpc.}
    \label{fig:r50tinf}
\end{figure}

\subsection{Present-day radial extent of the debris}

There is a growing body of evidence suggesting that the GE/GS is almost completely contained within a galactocentric radius of $\sim$25-30kpc, which coincides with the `break' radius of the stellar halo in the MW \citep{Deason2018, Lancaster2019}. A quick inspection of Fig.~\ref{fig:projs} suggests a wide range in the morphology and radial extension of the remnants for the S1-S4 satellite examples. In Fig.~\ref{fig:r50tinf}, we quantify the radial extent of the debris in our full sample of ancient radial mergers (ARMs, solid red points). For reference, we also show all satellites that have merged with our sample of $154$ disky MW-like centrals (semi-transparent symbols). The radial extent is characterized by $r_{*, \rm deb}$, the radius containing 50\% of the stars in each mergers' debris and symbols are color-coded by their median orbital anisotropy $<\beta>$.

There is a clear trend between the radial extent of the debris and its radial anisotropy: the more radial the orbit the more extended the debris.
This trend can be understood in terms of orbital dynamics, where at a fixed angular momentum (which is given by the orbit of the satellite), a circular orbit will minimize the radii \citep[see for instance Eq. 3.25b in ][]{Binney2008}. Large annisotropy values ($\beta > 0$) correspond to radially-biased orbits with large eccentricities, explaining the more extended distribution of stars in such orbits. 

In this context, the rather compact distribution inferred for GE/GS ($r < 30$ kpc) combined with its very large measured orbital anisotropy ($\beta \sim 0.95$) makes it a rather rare event. More than $80$\% of the simulated objects with infall times and orbital anisotropies similar to the GE/GS (labelled ARMs) have $r_{*, \rm deb}\geq45$ kpc, too large to be considered analogs of the observed GE/GS. We note that the average stellar mass radius of the disk-dominated MW analogs is $r_h \sim 10$ kpc, indicating that, typically, the debris of early radial mergers is expected to be quite extended, similar to those illustrated for S1 and S3 in Fig.~\ref{fig:projs}. 

Instead, GE/GS in the MW presents a compact morphology, perhaps more reminiscent of that of S2 in our sample. Therefore, in what follows we define a further subsample of our ARMs set by additionally requiring that their debris at $z=0$ fulfills the criterion $r_{*, \rm deb} < 25$ kpc. The resulting $6$ satellites have on average 80$\%$ of their stellar debris within $\sim$36 kpc, in better agreement with estimates for GE/GS. We hereafter refer to this subsample as compact ancient radial mergers (CARMs), highlighted with black stars in Fig.~\ref{fig:r50tinf}.

\begin{centering}
\begin{figure*}
	\includegraphics[width=2\columnwidth]{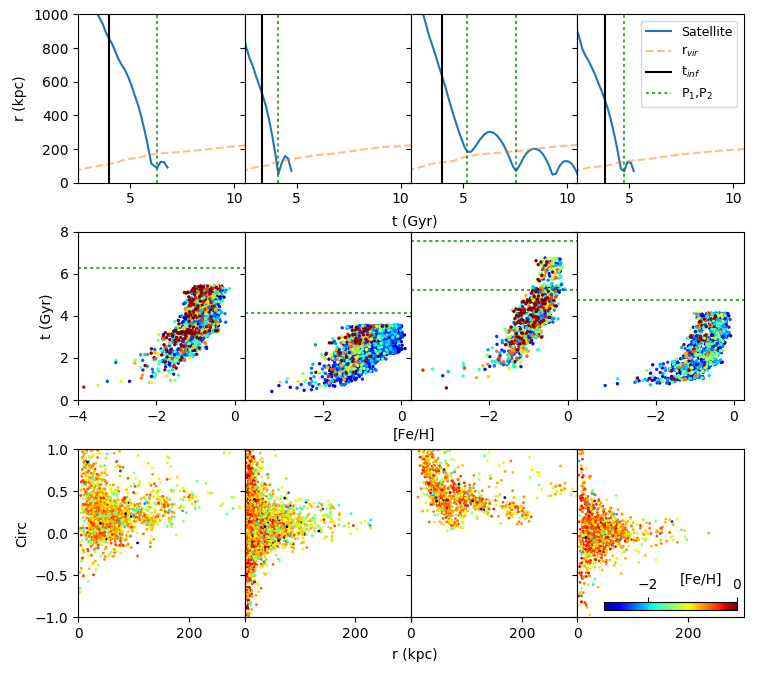}
    \caption{Orbits (top), metallicity (middle) and circularity (bottom) of stars deposited by ARMs examples S1-S4.
    \textit{Top row}: infall time is highlighted by black vertical lines while dashed orange lines indicate the virial radius evolution of their disk-dominated MW-like central. We also indicate the times of pericenter passages with dotted green lines. \textit{Middle row}: Time vs. metallicity of their stars, colored by galactocentric radius (same colorbar as Fig.~\ref{fig:encel}). Note that stars stop forming roughly at the time of their pericenters (green dotted lines). \textit{Bottom row}: Circularity of stellar debris as a function of galactocentric radius, colored by stellar metallicity as indicated by the color bar. S2 and S4 show a large fraction of radial, counterrotating orbits in agreement with GE/GS. Notice that the circularity does not appreciably change with radius and that no significant metallicity gradient with radius is found.}
    \label{fig:big}
\end{figure*}
\end{centering}

Besides the debris' radial extension, our numerical simulations give predictions of other dynamical and stellar properties for our sample of ARMs. Fig.~\ref{fig:big} further explores the orbits (top), metallicity-stellar age (middle) and circularity (bottom) at $z=0$ of the stars belonging to satellites S1-S4 (left to right). We use $[\alpha/\rm Fe] = 0.2$ to compute the simulations total metallicity $Z$ into the iron abundance [$\rm Fe/H$] following \citet{Salaris2005}. Stars are color coded according to their present-day distance, using the same scale than in Fig.~\ref{fig:encel}. 

\subsection{Age and metallicity of the debris}

The middle row in Fig.~\ref{fig:big} shows a wide range of metallicities associated to stars in a single satellite. This range overlaps well with the estimated metallicity of GE/GS, log([Fe/H]) $\sim -1.5$ \citep{Helmi2018}. We find no significant segregation with present-day distance of the stars (see color coding as in Fig.~\ref{fig:encel}), meaning that the remnants of GE/GS may be spotted outside of the solar neighborhood by looking at stars with similar ages and metallicities of the already-identified debris. However, we caution that the numerical resolution of Illustris and in particular the gravitational softening, $\epsilon_{*}=\epsilon_{\rm DM} \sim 0.7$ kpc, is comparable to the sizes of these satellites, which may be preventing us from resolving any population gradients within the satellites and driving the lack of correlation with present-day distance in this simulations.

Encouragingly, we detect a clear cut-off of the star formation associated to the infall of these objects. The age of the youngest stars is a very good indicator of the time of the first pericenter passage, as indicated by the green dotted lines in the top and middle rows. This provides partial validation to the observational interpretation of stellar age of the debris as the time of the merger. Previous works have placed the merger of GE/GS at $t\sim 6-10$ Gyr ago (see Sec.~\ref{sec:intro}). In general, our ARMs coalesce shortly after infall, although some exceptions showing several apocenters and pericenters may also be found (see for instance S3, which coalesces after the second pericenter). The time when the subhalo merges (the end of the orbits in the top row) is always later than the cut off in ages of the stars seen in the plots, confirming that the satellites are indeed quenched before being disrupted. 

\subsection{Orbit and rotation of the debris}

An interesting feature of the GE/GS event is the large fraction of counterrotating stars with respect to the rotation of the MW's disk. In our simulations we find that counterrotating orbits are not difficult to obtain, although are perhaps not the most likely. Examples of this can be seen in the bottom row of Fig.~\ref{fig:big} which shows, from left to right, the present-day circularity vs. galactocentric radius of the stars in S1-S4 debris. 

Stars from S1 and S3 have almost completely positive circularities (i.e. are prograde), while those from S2 and S4 have a significant retrograde component (f$\rm_{cntr}=32\%$ and 49$\%$ respectively). 
In fact, out of our whole ARMs sample of $37$ objects, the median fraction of stars in the debris that are retrograde is 39$\%$, indicating that the large fraction of counterrotating stars in GE/GS is not difficult to accommodate within the $\Lambda$CDM assembly of MW-like galaxies. 

\begin{figure}
	\includegraphics[width=\columnwidth]{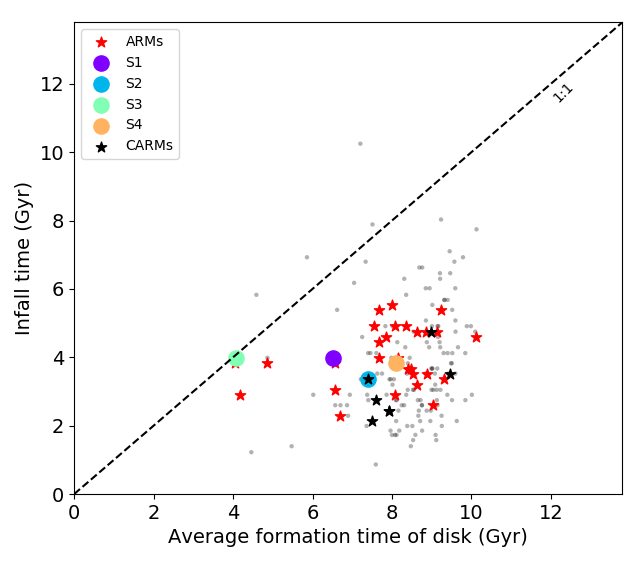}
    \caption{Infall time of satellites vs. (averaged) formation time of the disks for our disk-dominated centrals with at least 1 ARM. Gray, red-starred and black-starred symbols show all satellites, ARMs and CARMs, respectively. Large colored circles show S1-S4 as before.  All centrals, with the exception of one, form their disks after the infall time of the ARM, explaining how the disk of the MW could have survived the merger with GE/GS.}
    \label{fig:diskage}
\end{figure}

\begin{centering}
\begin{figure*}
	\includegraphics[width=2\columnwidth]{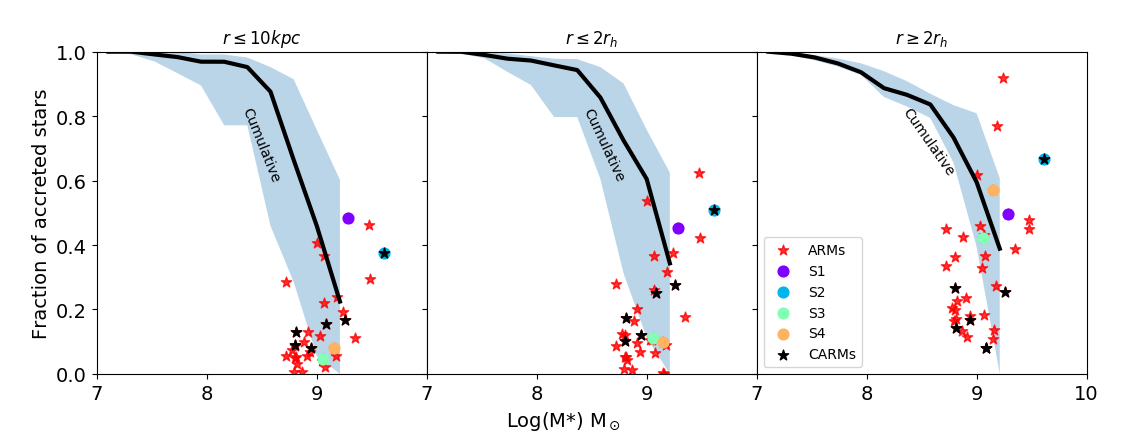}
    \caption{Stellar halo fraction contributed by our identified ARMs (red starred), CARMs (black starred) and S1-S4 (color circles). From left to right, we differentiate the very inner halo ($r<10$ kpc), inner halo ($r<2r_{h,*}$) and outer halo ($r>2r_{h,*}$), respectively. 
    Although some of these individually comprise up to $50\%$ of the inner and outer components by themselves, in general GE/GS-like events contribute a much smaller fraction of the accreted halo. For instance, ARMs and CARMs individually contribute a median of 12$\%$ and 21$\%$, respectively, suggesting that a larger number of relatively massive events is the norm for these MW-like analogs. This is confirmed by the median cumulative stellar halo fraction deposited by satellites above a given $M_*$ (solid black curve), with shaded blue regions indicating the $25\%$-$75\%$ quartiles in our simulated sample. Note that S2 is the only CARM that contributes significantly to the inner halo, as is inferred for GE/GS. Interestingly, S2 also shows a large contribution to the outer halo, inviting further exploration of the outer halo in MW data.}
    \label{fig:fracinout}
\end{figure*}
\end{centering}

In the classical view where disk galaxies inhabit halos with enough angular momentum \citep{Frenk1991, Mo1998} and where angular momentum has been preserved and coherently added over time \citep{Sales2012, GarrisonKimmel2018}, the presence of counterrotating debris in disky MW analogs is somewhat unexpected. However, this may be better understood when considering the early time of the GE/GS merger event. Fig.\ref{fig:diskage} shows the infall time of satellites for our disky MW sample, compared to the median formation time of the disk in each central. The disk formation time is computed as the median age of stars kinematically associated to the disk, defined with the criteria: r$_{xy}<2r_h$, |z|<3 kpc, circ>0.5 . 

The vast majority of cases lay to the right of the 1:1 line (blue dashed), indicating that the infall times of satellites occur well before the typical epoch of formation of the disk. In particular, our set of most compact radial merger events (or CARMs, black starred symbols) and closest GE/GS analogs, have infall times $t_{\rm inf} \sim 4$ Gyr, about half the typical age of disk formation, $t \sim 8$ Gyr. It is therefore possible that the orientation of early merger events like GE/GS may be considered random with respect to the later established direction of disk' rotation, helping explain the large fraction of counterrotating stars found in these remnants. 

\section{The contribution of GE/GS to the stellar halo build up}
\label{sec:halo}
\begin{figure}
	\includegraphics[width=\columnwidth]{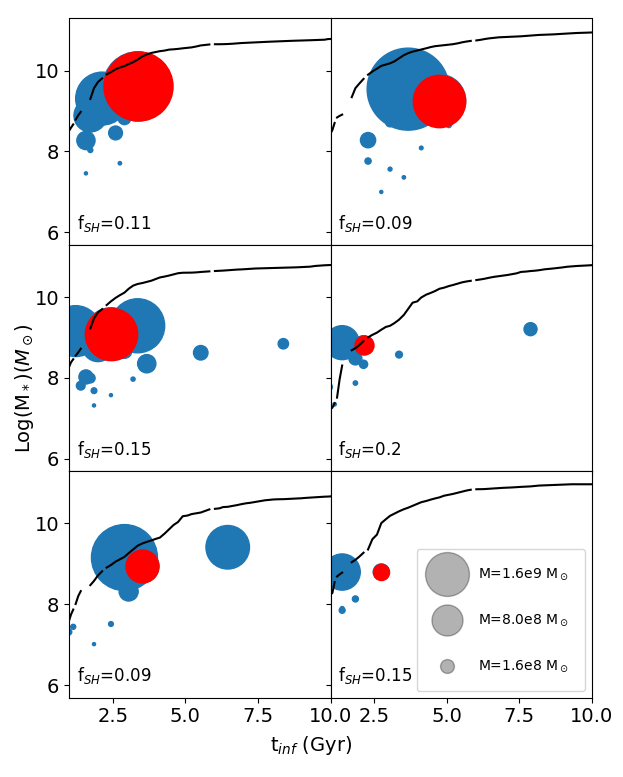}
    \caption{Building blocks of the accreted stellar halos in disk-dominated MW-analogs that have experienced a CARM. The figure shows the maximum stellar mass vs. infall time of the merged satellites in these centrals. 
     The size of each point scales with the number of stars contributed to the inner stellar halo by each merger event. Red circles highlight the CARM contribution. Most of the accreted stars in the galaxy are acquired early on and, with the exception of S2 (top left panel) the debris of the CARM does not dominate the accreted material. Solid black line indicates for comparison the stellar mass of the central galaxy as a function of time and the stellar halo fraction is quoted for each central.}
    \label{fig:circles}
\end{figure}

Stars now associated to the GE/GS debris dominate the local sample of stellar halo stars in the solar neighborhood \citep{Helmi2018, Belokurov2018}. However, we have fewer constraints on the overall contribution of the GE/GS event to the build up of the total galaxy-wide stellar halo of the MW. \citet{Deason2018} have linked the $r\sim 30$ kpc break in the MW's stellar halo to the pileup of stars at the apocenters of the GE/GS orbit, implying that the majority of the inner halo in the MW was built by the single merger with the GE/GS. It is important to bear in mind that multiple shells possibly associated to previous apocenter passages may also be expected from the debris. 

More quantitatively, orbital modeling of the stars in the GS debris estimate that up to 50\% of the stellar halo stars within $r\leq25$ kpc were brought in by the GS event alone \citep[see e.g., ][]{Lancaster2019, Iorio2018}. We can use our simulations to shed light on the expected contribution of GE/GS events to the stellar halo and aid the interpretation of solar neighborhood-based results in a galaxy-wide context. Moreover, we can also find clues as to the contribution of the GE/GS event to the outer halo and guide future identification of this debris in the outer MW regions. 

We restrict our study to the $37$ disk-dominated MW galaxies that have experienced at least one ARM-like merger and identify their accreted stellar component within the stellar halo. We further divide the stellar halos in 3 regions: very inner accreted halo ($r<10$ kpc), inner halo $r<2 r_h$ with $r_h$ the half mass radius of the stars in each host, and outer halo, $r>2 r_h$. The solid black curves in Fig.~\ref{fig:fracinout} show, for these three regions, the median cumulative fraction of the stars in the accreted halo that were contributed by any satellite of a given (maximum) stellar mass or above (x-axis). The $25\%$-$75\%$ quartiles are also highlighted in blue shading. We find that the largest contributors to the build up of their stellar halos are satellites in the mass range $M_*\sim 5 \times 10^8$-$5 \times 10^9\; \rm M_\odot$. Furthermore, our simulations predict no significant differences between the inner vs. outer regions, except for a slightly enhanced contribution of low mass mergers to the outer $r>2r_h$ stellar halos compared to the inner regions.  

To evaluate the individual contribution of our identified GE/GS events, we overplot in Fig.~\ref{fig:fracinout} the (non-cumulative) fraction of stars deposited by each identified old and radial merger (ARMs, red stars). Surprisingly, we find a rather modest contribution, typically accounting for $10\%$ of the accreted inner halo and up to $20\%$ for the outer regions. Even selecting those that are the most compact events (CARMs, black stars) as closest analogs to GE/GS suggests an average contribution of less than 20\% for $r<10$ kpc and as much as 25\% of the outer accreted halo. We conclude that, for this sample of MW-like analogs, GE/GS-like mergers seem to provide a more minor contribution to the stellar halo than that inferred for the case of the MW. Our theoretical results are therefore better aligned with other estimates that place the GS as a non-dominant contributor \citep[perhaps lower end of the modeling in ][]{Helmi2018, Mackereth2019b}.

Interestingly, some exceptions occur, for instance, such as our selected S2 satellite, whose stars alone account for $\sim 50$\% of the inner stellar halo in this central. Encouragingly, S2 also satisfies the compactness CARMs criteria (cyan/black symbol) and seems to be our best  match to the observed GE/GS in the MW. 

If the GE/GS analogs identified in Illustris are mostly minor contributors to the stellar halo of their hosts, one might wonder what else helps build the stellar halos in such centrals. We explore this in Fig.~\ref{fig:circles}. Each circle is a merger experienced by a central galaxy that has experienced a CARM (shown in red). The sizes of the circles are proportional to the number of stars deposited in the inner stellar halo and the black line shows the stellar mass of the central. The top left panel presents the CARM with the largest contribution to the inner accreted stellar halo, which is satellite S2. For this central, the CARM is the latest merger and also the largest contributor, in good agreement with estimates of GE/GS in the MW. However, both Fig.~\ref{fig:fracinout} and Fig.\ref{fig:circles} show this merger is an anomaly rather than the norm. Yet, it is still plausible within the wide diversity of stellar halos predicted by $\Lambda$CDM.

The right-most panel in Fig.\ref{fig:circles} may be used to guide future searches for the remaining GE/GS debris. Our simulations indicate that the most compact of our GE/GS analogs also build up to $\sim 20\%$ of the stars in the outer stellar halo. Moreover, our closest analog S2 provides $60\%$ of the accreted outer halo in this central, suggesting the idea that a sizable fraction of the MW's outer halo could be associated to the GE/GS event. Although the peculiarity of this event prevents us from having a statistical basis to make predictions about the singular case of the MW, our results support a scenario where more of the GE/GS remnant might be uncovered in the future by studying not only further into the inner regions of the MW's stellar halo, but also perhaps outwards of the solar region, where long dynamical times may be more favorable to the identification of merger debris.

\section{Summary and Discussion}
\label{sec:concl}

We use a sample of MW analogs identified in the Illustris simulations to study the present-day remnants of mergers comparable to the Gaia Enceladus/Sausage (GE/GS) event. Observationally, the GE/GS object has been inferred to be old $t_{\rm inf}\sim10$ Gyr ago, in a radial orbit (orbital anisotropy $\beta=0.8-0.9$) and massive ($M_*\sim10^9 M_\odot$). It largely dominates the stellar halo stars in the solar neighborhood \citep{Helmi2018} and is thought to make up to $\sim50\%$ of the inner stellar halo in our Galaxy \citep{Lancaster2019}. These conclusions, however, originate from a small and local fraction of its inferred mass, where the high quality of the data allows for the identification of substructure in position and velocity space. Much is unknown about the distribution, kinematics and stellar properties of the rest of the mass estimated for the progenitor of GE/GS. The large volume of Illustris allows for a statistical basis to place GE/GS-like events within the cosmological predictions of the $\Lambda$CDM model and to make statistical predictions about the remaining stream.  

From a sample of 1115 isolated galaxies in the mass range $M_{200} = 0.8$-$2.0 \times 10^{12}\; \rm M_\odot$, we select $154$ Milky Way analogs that show a disk-dominated morphology. In agreement with the early accretion inferred for the merger of GE/GS, our sample of disky MW-like hosts naturally shows an early assembly history, with $86\%$ of the mergers with satellites of stellar mass $M_* > 5 \times 10^8\; \rm M_\odot$ happening by $t<6$ Gyr ($z\sim0.9$). Moreover, almost all centrals in our sample (80\%) show at least $1$ merger with a satellite of such mass. The stellar content of GE/GS and its early accretion seem to be a natural prediction for our Galaxy within the cosmological model. 

We focus our analysis on the study of early mergers ($t<8$ Gyr ago) in these centrals that involved massive ($5 \times 10^8\leq M_* \leq 5 \times 10^9 M_\odot$) satellites with a stellar debris characterized (today) by radial orbits ($\beta >0.5$). We find that $37$ ($\sim 25\%$) of our centrals have had at least one such ancient radial merger (ARM). Despite the specific constraints on the identification of these ARMs, there is a wide range of morphologies and kinematics associated to the stellar streams and remnants of these events. Looking in the space of velocities and/or energies, as GE/GS was identified, we find that good GE candidates also seem to be consistent with GS, providing support for a scenario where both structures could be, predominantly, the same. 

Two remaining properties have been highlighted from observations of GE/GS. First, a significant counter-rotating component. Second, it appears likely to have deposited most of the stars within $25$-$30$ kpc, as inferred by the break in the stellar halo density profile \citep{Deason2018, Lancaster2019}. We find that counterrotation is rather common in our sample of ARMs, with 43$\%$ of satellites depositing at least $40\%$ of their stars in present-day counter-rotating motion. The early times of these mergers, coupled to a later build up of the disk in the simulations help explain the large number of stars in the debris that are counterrotating. 

Compactness of GE/GS debris in our sample is significantly more rare given its radial orbit. Massive and early mergers in Illustris with $\beta>0.5$ show median radius containing half of the mass $r_{*,\rm deb} \sim 45$ kpc; with the most extreme objects extending to $r_{*,\rm deb} \sim 143$ kpc, well into the dark matter halos of their host galaxies. In general, the more radial the orbit, the more extended the debris. Instead, studies of GE/GS place it mostly within $30$ kpc despite its $\beta \sim 0.9$. Only $6$ ($\sim 16\%$) of our radial mergers have a comparably compact radial extension today, with $r_{*,\rm deb} < 25$ kpc.  We refer to them as compact ARMs, or CARMs for short. 

We can use our simulations to shed light on the contribution of GE/GS to the build up of the global (i.e. beyond the solar neighborhod) stellar halo in the MW. Considering our $37$ disk-dominated MW-like centrals, we find that their accreted components are built generally by a few (but more than one) relatively massive accretion events. Individual objects are unlikely to dominate the entire stellar halo. For instance, ARMs contribute only $\sim 9\%$ (median) of the inner stellar halo within 10 kpc and 12$\%$ within $\sim 25$ kpc (corresponding to twice the average half mass radius of the centrals). These numbers increase to $14\%$-$21\%$ when considering the more compact CARMs. In Illustris, for those MW-like analogs that have experienced a GE/GS like event (defined as ancient, massive, radial and compact), it is not a single event but the contribution of $2$-$3$ $M_* > 5 \times 10^8\; \rm M_\odot$ satellites that make up to $90\%$ of the stars in the inner halos. 

There are, however, a few extreme cases where we find ARMs and CARM events contributing up to $\sim 60\%$ of the accreted inner halo on an individual basis. This is more in line with some results that place GS/GE as the dominant builder of the MW's inner halo \citep{ Belokurov2018, Deason2018}, although different estimates suggest a more modest contribution \citep{Helmi2018,Mackereth2019b}. In our sample, we find one good GE/GS analog (named S2 throughout the paper) that shows a compact enough distribution to be comparable to GE/GS and that simultaneously brought in $50\%$ of the inner accreted halo. 

Interestingly, for the same particular host galaxy, the merger event S2 also contributes significantly to the outer stellar halo, perhaps suggesting that hidden stars of GE/GS lie outside of r>30 kpc, waiting to be discovered. The predictions for the amount and distribution of such outer halo stars vary among our 6 identified CARMs, with a median of $\sim 3.25 \times 10^8\; \rm M_\odot$ outside $\sim 25$ kpc, but as much as $1.5 \times 10^9\; \rm M_\odot$ for the most promising case of S2 in our sample. The median radius for these outer stars is $\sim 40$ kpc, but they can extend as far as $\sim 230$ kpc. 

If GE/GS stars could be found beyond $r \sim 25$ kpc, our simulations predict that their age and metallicities should be comparable to the section already identified of GS/GE, a conclusion that needs confirmation from higher resolution experiments and a more detailed ISM treatment than in our simulations (along the lines of work proposed by \citet{Bignone2019}). Our sample has more predictive power for dynamical quantities instead. We find that in all our CARMs the stellar debris contributing to the outer stellar halo preserves a similar radial orbit distribution as the stars deposited in the inner regions of the disk. For the specific case of our best analog S2, we find a moderate evolution of the orbit orientation, such that the fraction of counterrotating stars in the outer halo is smaller ($\sim 26\%$) than in the inner halo ($\sim 42\%$). 

It is unclear whether significant amounts of stars belonging to GE/GS could be hidden in the outer halo of the MW, but theoretical predictions strongly support such a case. The data shows a clear drop in the number density of stars kinematically associated to GE/GS beyond $r \sim 30$ kpc \citep{Deason2018,Lancaster2019}. However, our study of $\sim 1000$ MW-mass galaxies indicates that even the most compact, early and massive mergers with comparable radial orbits than GE/GS deposit roughly $0.13$-$1.7 \times 10^9\; \rm M_\odot$ in the outer stellar halo. It is then possible (and even likely) that a significant fraction of the GE/GS progenitor is in a more diffuse stream extending into the outer realms of the Milky Way. Future observational efforts targeting the oldest and most radially-biased stars may be able to recover the earliest stripped shells from the GE/GS progenitor.

\section*{Acknowledgements}

We would like to acknowledge the useful discussions we had with V. Belokurov, A. Fattahi, A. Deason, C. Frenk and R. D'Souza regarding this work. LME and LVS acknowledge support from the Hellman Fellowship. 






\bibliographystyle{mnras}
\bibliography{master} 



\bsp	
\label{lastpage}
\end{document}